\documentclass[pdflatex,sn-mathphys-num]{sn-jnl}

\usepackage{graphicx}%
\usepackage{multirow}%
\usepackage{amsmath,amssymb,amsfonts}%
\usepackage{amsthm}%
\usepackage{mathrsfs}%
\usepackage[title]{appendix}%
\usepackage{xcolor}%
\usepackage{textcomp}%
\usepackage{manyfoot}%
\usepackage{booktabs}%
\usepackage{algorithm}%
\usepackage{algorithmicx}%
\usepackage{algpseudocode}%
\usepackage{listings}%

\theoremstyle{thmstyleone}%
\theoremstyle{thmstyletwo}%

\theoremstyle{thmstylethree}%

\def\usepng{1}  

\begin{document}

\title[SASVi]{SASVi - Segment Any Surgical Video}

\author*[1]{\fnm{Ssharvien} \sur{Kumar Sivakumar}}\email{ssharvien.kumar.sivakumar@gris.tu-darmstadt.de}
\equalcont{These authors contributed equally to this work.}

\author*[1,2]{\fnm{Yannik} \sur{Frisch}}\email{yannik.frisch@gris.tu-darmstadt.de}
\equalcont{These authors contributed equally to this work.}

\author[1]{\fnm{Amin} \sur{Ranem}}

\author[1]{\fnm{Anirban} \sur{Mukhopadhyay}}

\affil[1]{\orgdiv{GRIS}, \orgname{TU Darmstadt}, \orgaddress{\street{Fraunhoferstr. 5}, \city{Darmstadt}, \postcode{64283}, \country{Germany}}}
\affil[2]{\orgdiv{NRAD}, \orgname{UM Mainz}, \orgaddress{\street{Langenbeckstr. 1}, \city{Mainz}, \postcode{55131}, \country{Germany}}}

\abstract{
\textbf{Purpose:} Foundation models, trained on multitudes of public datasets, often require additional fine-tuning or re-prompting mechanisms to be applied to visually distinct target domains such as surgical videos. Further, without domain knowledge, they cannot model the specific semantics of the target domain. Hence, when applied to surgical video segmentation, they fail to generalise to sections where previously tracked objects leave the scene or new objects enter.
 
\textbf{Methods:} We propose \emph{SASVi}, a novel re-prompting mechanism based on a frame-wise object detection \emph{Overseer} model, which is trained on a minimal amount of scarcely available annotations for the target domain. This model automatically re-prompts the foundation model \emph{SAM2} when the scene constellation changes, allowing for temporally smooth and complete segmentation of full surgical videos. 
 
\textbf{Results:} Re-prompting based on our \emph{Overseer} model significantly improves the temporal consistency of surgical video segmentation compared to similar prompting techniques and especially frame-wise segmentation, which neglects temporal information, by at least 2.4\%. Our proposed approach allows us to successfully deploy \emph{SAM2} to surgical videos, which we quantitatively and qualitatively demonstrate for three different cholecystectomy and cataract surgery datasets.
 
\textbf{Conclusion:} \emph{SASVi} can serve as a new baseline for smooth and temporally consistent segmentation of surgical videos with scarcely available annotation data. Our method allows us to leverage scarce annotations and obtain complete annotations for full videos of the large-scale counterpart datasets. We make those annotations publicly available, providing extensive annotation data for the future development of surgical data science models.
}




\maketitle

\section{Introduction}
\label{sec:intro}

Surgical video segmentation is crucial in advancing computer-assisted surgery, aiding intraoperative guidance and postoperative assessment. However, modern Deep Learning (DL) solutions require large-scale annotated datasets to be effectively trained. Gathering \textbf{annotations} in the form of \textbf{complete segmentation masks} requires substantial effort since creating full per-pixel annotations is a highly tedious task \cite{sanner2024detection}. This issue is multiplied in surgical process modelling, where DL solutions are often targeted at analysing long video sequences \cite{al2019cataracts,twinanda2016endonet}, significantly increasing the annotation effort along the temporal axis.

Large \textbf{foundation models} have lately emerged, trained on multitudes of publicly available large-scale datasets and often multiple tasks in parallel. These methods have proven to be successful when applied out of the box or fine-tuned to other domains \cite{ma2024segment,yu2024sam,chen2024sam2}. Yet, their application for computer-assisted surgery is either limited to frame-wise segmentation without incorporating temporal information \cite{sheng2024surgical,yue2024surgicalsam,chen2024sam2}, tracking only single tool classes \cite{wu2024real,liu2024surgical} or relying on manual prompting \cite{lou2024zero,yu2024sam}. 

\begin{figure}[htbp]
    \centering
    \if\usepng1
        \includegraphics[width=\textwidth]{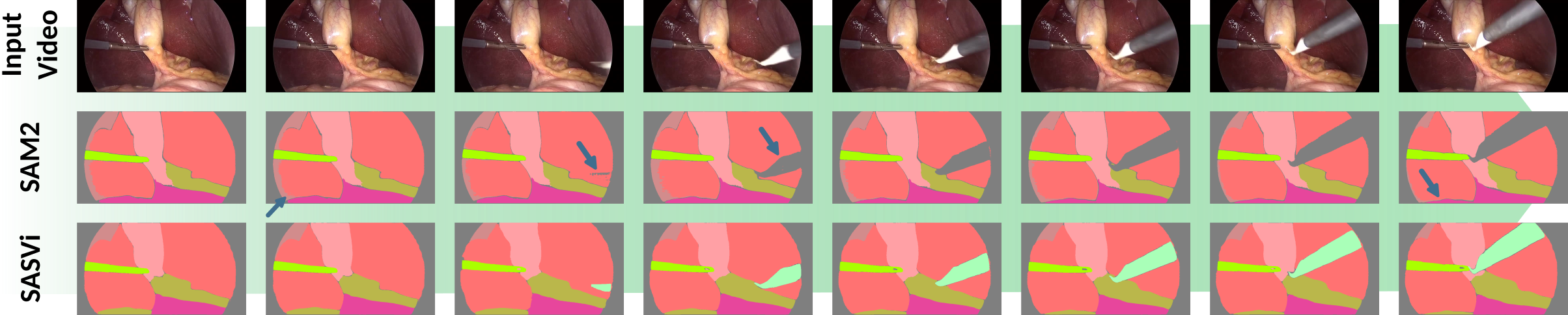}
    \else
        \includegraphics[width=\textwidth]{figures/introduction_v3.eps}
    \fi
    \caption{\textbf{SAM2 Failure Case.} Video segmentation with \emph{SAM2} struggles with objects leaving or entering the scene (middle row; the \emph{electrocautery} is missed and predicted as background). \emph{SASVi} mitigates this issue by leveraging a frame-wise overseer model, producing temporally smooth and complete segmentations from scarce annotation data (bottom row).}
    \label{fig:intro}
\end{figure}

\emph{SAM2} \cite{ravi2024sam} recently emerged as a robust video object tracking and segmentation tool but still relies on \textbf{manual prompting} and can fail to generalise to \textbf{video sections where entities leave the scene or new objects enter}, as visualised in Figure \ref{fig:intro}. Such events happen frequently in surgical video data when other instruments are used in subsequent surgical phases or when the camera moves during laparoscopy. Usually, such moments would require a re-prompting of the new entities to track, again increasing the manual effort of the clinician or machine learning engineer in the loop \cite{wang2023sam}. Further, without external domain knowledge, the method does not model the semantic meanings of tracked entities, rather than just performing consistent segmentation of tracked objects throughout a video.

We propose \emph{Segment Any Surgical Video (SASVi)}, a novel video segmentation pipeline including a re-prompting mechanism based on a supportive frame-wise overseer model which runs in parallel to \emph{SAM2}. Precisely, we deploy an object detection model, pre-trained on small-scale surgical segmentation datasets, to monitor the entities currently present in the video. The dual nature of models such as \textit{Mask R-CNN} \cite{he2017mask}, \emph{DETR} \cite{carion2020end} or \emph{Mask2Former} \cite{cheng2022masked} allows us to rely on the object detection part of the model to detect when untracked classes enter the scene or previously tracked entities leave. We can then intercept such time points and use the model's segmentation part to segment the current frame. The obtained segmentation mask is then used to sample new prompting anchors for each currently present entity, including their semantic meaning. These anchor prompts are subsequently utilised to re-prompt \emph{SAM2}, which then continues the segmentation.

With this re-prompting of our overseer model, trained on scarcely available annotations, we can successfully leverage \emph{SAM2}'s excellent temporal properties to segment long video sequences of various surgical modalities with limited available annotation data. We quantitatively and qualitatively demonstrate on three prominent cholecystectomy and cataract surgery datasets that our method generates temporally smooth and consistent semantic segmentations of complete surgical video sequences. This further allows us to provide complete segmentation annotations of large-scale surgical video datasets for the public without additional manual annotation effort. 

\paragraph{Contributions}
\begin{itemize}
    \item We are the first to propose an automated re-prompting mechanism based on an object detector for deploying \emph{SAM2} for temporally smooth and consistent semantic segmentation of arbitrary surgical video domains with scarce annotation data.
    \item We deploy our method to leverage small-scale annotated surgical segmentation datasets into fully annotated publicly available large-scale segmentation annotations of their origin videos, demonstrated for the cholecystectomy dataset \emph{Cholec80} and the cataract surgery datasets \emph{Cataract1k} and  \emph{CATARACTS}. 
\end{itemize}

\section{Related Work}
\label{sec:rw}

For \textbf{segmenting surgical videos}, Wang et al. \cite{wang2021efficient} have introduced a dual-memory network to relate local temporal knowledge with global semantic information by incorporating an active learning strategy. Zhao et al. \cite{zhao2021anchor} combine meta-learning with anchor-guided online adaption to improve domain transfer generalisation. COWAL \cite{wu2024correlation} deploys an active learning strategy based on model uncertainty and temporal information to improve video segmentation. However, these approaches require access to large-scale annotated data for their specific target or visually similar source domains.

\textbf{Foundation models}, trained on large-scale computer vision datasets, have been successfully deployed in the recent past to demonstrate generalisation capabilities for segmentation \cite{kirillov2023segment}. This model has found a wide range of applications in medical imaging \cite{ma2024segment,ranem2024uncle}.

In the \textbf{surgical context}, \emph{SurgicalSAM} \cite{yue2024surgicalsam} eliminates the need for explicitly prompting \emph{SAM}\cite{kirillov2023segment} by introducing a prompt encoder that generates prompt embeddings automatically, alongside contrastive prototype learning to distinguish visually similar tools better. \emph{Surgical-DeSAM} \cite{sheng2024surgical} combines \emph{SAM} with a \emph{DETR} model for tool detection and re-prompts SAM using bounding boxes, enabling multi-class segmentation. While these approaches improve frame-wise segmentation, they do not leverage temporal information from videos.

The \emph{Segment Anything Model 2 (SAM2)} \cite{ravi2024sam} extends \emph{SAM} \cite{kirillov2023segment} for \textbf{video segmentation}. It achieves temporally smooth segmentations by introducing a memory buffer of previous information. \emph{SAM2-Adapter} \cite{chen2024sam2} extends \emph{SAM2} by introducing trainable adapter layers to incorporate task-specific knowledge and has been successfully applied to frame-wise polyp segmentation. \emph{Surgical SAM2} \cite{liu2024surgical} implements a frame-pruning mechanism to reduce memory and computation costs, addressing challenges associated with processing long sequences of surgical video frames. Yu et al. \cite{yu2024sam} evaluate \emph{SAM2} on surgical videos using manual point and box prompts. They observe robust results but also point to the method's limitations when dealing with synthetic data, where performance degrades due to image corruptions and perturbations. Similarly, zero-shot segmentation using SAM2 has been explored for surgical tool tracking in endoscopy and microscopy data, proving effective for multi-class tool segmentation \cite{lou2024zero}. However, unlike our proposed approach, these methods still rely heavily on manual prompting and do not implement re-prompting mechanisms, hence suffering from performance decreases when entities leave or enter the scene.
\section{Method}
\label{sec:meth}

This section outlines the components of our approach, \emph{SAM2} and the \emph{Overseer} model, before describing our inference pipeline for video segmentation.

\subsection{SAM2: Segment Anything in Images and Videos}
Given a video sequence $V := \{v_t\}_{t=1}^{T}, v_t \in \mathbb{R}^{3 \times H \times W}$, the \emph{SAM2} model $F(v)$ encodes the first frame $v_1$ into a latent representation by a hierarchical \emph{image encoder} network. Various prompts in the form of anchor points, bounding boxes or segmentation masks are equally encoded by a \emph{prompt encoder}. Both representations are then fed into the model's \emph{mask decoder} to produce the segmentation mask $\Bar{m}_1$, which is then again encoded by the \emph{memory encoder}. Encoded masks and frames are added to a \emph{memory bank}. For subsequent frames $v_t$ of the sequence $V$, entries from that memory bank are conditioning the current frame encoding in a \emph{memory attention} module before feeding it into the \emph{mask decoder} to predict $\Bar{m}_t$. We refer to Ravi et al. \cite{ravi2024sam} for further details.

\begin{figure}[htbp]
    \centering
    \if\usepng1
        \includegraphics[width=\textwidth]{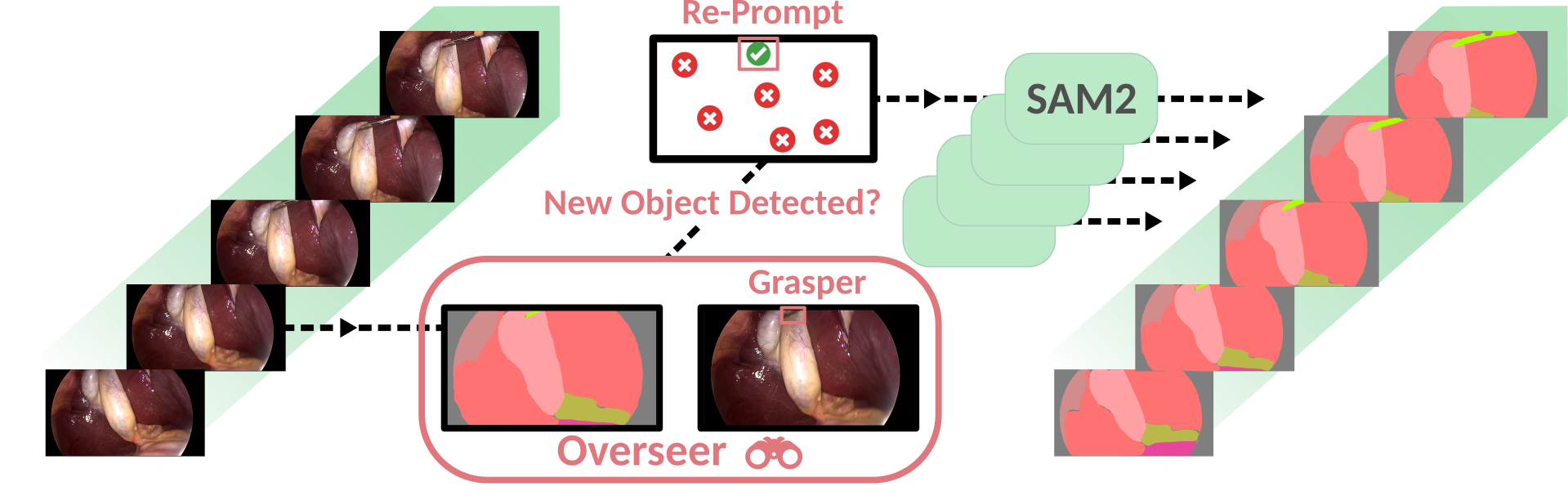}
    \else
        \includegraphics[width=\textwidth]{figures/inference_scheme_v2.eps}
    \fi
    \caption{\textbf{SASVi Inference Scheme.} Our frame-wise \emph{Overseer} model (\includegraphics[height=0.25cm]{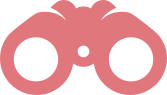}) captures time points at which previously untracked entities enter the scene or tracked objects leave. At that moment, it re-prompts \emph{SAM2} with predictions from that frame.}
    \label{fig:sasvi}
\end{figure}

\subsection{Object Detection Overseer Model}
To serve as an \emph{Overseer} model for \emph{SAM2} \cite{ravi2024sam}, we pre-train \emph{Mask R-CNN} \cite{he2017mask}, \emph{DETR} \cite{carion2020end} and \emph{Mask2Former} \cite{cheng2022masked} on the scarcely annotated datasets. Given an image frame $v_t$, the methods' \emph{Region Proposal Network} (RPN) predicts \emph{Regions of Interest} (ROIs), from which the \emph{Object Detection Stream} predicts bounding boxes $t := (x_\text{min}, x_\text{max}, y_\text{min}, y_\text{max}) \in [0,1]^{N_\text{bb} \times 4}$ for $N_\text{bb}$ objects and class probabilities $p \in [0,1]^{N_\text{cls} \times C}$ for $N_\text{cls}$ objects and the $C$ classes of the dataset. In parallel, the models' \emph{Segmentation Stream} predicts probability masks $m \in [0,1]^{N_\text{mask} \times H' \times W'}$ for $N_\text{mask}$ objects, where $(H',W')$ are the ROI dimensions. Example predictions of both streams of \emph{Mask R-CNN} are visualised in Figure \ref{fig:frame_qual}.

The models are trained by minimising
\begin{equation}
    \mathcal{L} = \frac{1}{N_\text{cls}} \sum_{i=1}^{N_\text{cls}} \mathcal{L}_\text{cls}(i) + \frac{1}{N_\text{bb}} \sum_{i=1}^{N_\text{bb}} \mathcal{L}_\text{box}(i) + \frac{1}{N_\text{mask}} \sum_{i=1}^{N_\text{mask}} \mathcal{L}_\text{mask}(i)
\end{equation}
with
\begin{align}
    \mathcal{L}_\text{cls}(i) &=  - \sum_{k=1}^C c^*_{ik}\log(p_{ik}) \text{,}\quad \mathcal{L}_\text{box}(i) = \text{smooth}{L_1}(t_i - t^*_i) \quad\text{and}\\
    \mathcal{L}_\text{mask}(i) &=  \frac{1}{H \times W} \sum_{x=1,y=1}^{H,W} - [m^*_{c^*_i,x,y}\log(m_{c^*_i,x,y}) + (1-m^*_{c^*_i,x,y} ) \log(1-m_{c^*_i,x,y})]
\end{align}

where $c^*$, $t^*$ and $m^*$ are the ground-truth class probabilities, bounding box coordinates and segmentation masks, respectively.

Unlike traditional segmentation models, our \emph{Overseers} can catch new instances of the same class, which the former would predict in a single mask. As further analysed in Supplementary Section \ref{sec:app_compute}, their lightweight design allows for efficient monitoring of the surgical videos in parallel to \emph{SAM2}.

\subsection{Segment Any Surgical Video}
Given a video sequence $V$, our method operates as follows:

In the initial frame $v_{t=1}$, we query the pre-trained \emph{Overseer} model $M(v)$ to predict a segmentation mask $m_{t=1} = M(v_{t=1})$.
Given this prediction, we store the current entities in a buffer as $B := \{c_1\}$, where $c_1 \leq C$ are the currently predicted classes.
The mask is used to prompt the \emph{SAM2} model $F(v_{t=1},m_{t=1})$, predicting the segmentation mask $\Bar{m}_{t=1}$. 
Subsequent frames $\{v_t\}_{t=2}^{T}$ are equally segmented with $F(v_t)$, producing temporally smooth segmentations.
In parallel, the \emph{Overseer} $M(v_t)$ predicts the classes $c_t$ and adds them to the buffer $B$. 

Once we reach a frame $v_t'$ where the class predictions in $B$ changed for more than $n_t$ time-steps, we perform the following: We track back the time point $t'-n_t$ where the change in classes first happened. We then sample anchor prompting points $a_{t'-n_t}$ from the \emph{Overseer} mask $m_{t'-n_t}$ and use these prompts in conjunction with mask $m_{t'-n_t}$ to continue the segmentation from that point in time. The threshold $n_t$ is introduced to minimise the impact of wrong predictions from $M(v_t)$ and is empirically set to $n_t = 4$. Further, the temporal back-tracking allows for correcting potential mistakes from $F(v)$ in the last $n_t$ time steps, smoothing out the predictions. This process is repeated until the full video $V$ is segmented as $\Bar{M} := \{\Bar{m}_t\}_{t=1}^T$.

The overall inference process is visualised in Figure \ref{fig:sasvi} and summarised as a pseudo-code formulation in Algorithm \ref{alg:pseudocode}.

\begin{algorithm}[htbp]
    \caption{\textbf{SASVi Inference Pseudocode.}}
    \label{alg:pseudocode}
    \begin{algorithmic}
        \Require Pre-trained \emph{Overseer} model $M(v_t)$, \emph{SAM2} model $F(v_t,a_t)$, surgical video sequence $\{v_t\}_{t=1}^T$, temporal buffer $B$ of size $n_t \geq 1$, anchor sampling size $n_a \geq 1$
        \State $m_1,c_1 \gets M(v_t)$ \textit{ // Predict the first frame using the Overseer.}
        \State $B \gets \{c_1\}$
        \State $\Bar{m}_1 \gets F(v_1,m_1)$ \textit{ // Prompt SAM2 with the predicted mask.}
        \State $t \gets 2$
        \While{$t \leq T$}
        \State $m_t, c_t \gets M(v_{t})$ \textit{ // Predict the current frame using the Overseer.}
        \State $B \gets B + \{c_t\}$
        \If{$t - n_t \geq 0 \text{ and new class in all of } B$}
            \State $a_{t-n_t} \gets \text{sample}(m_{t-n_t}, n_a)$ \textit{ // Sample anchor points for new entity.}
            \State $\Bar{m}_{t-n_t} \gets F(v_{t-n_t}, a_{t-n_t}, m_{t-n_t})$ \textit{ // Re-prompt SAM2.}
            \State $t \gets t-n_t+1$
        \Else
            \State $\Bar{m}_{t} \gets F(v_{t})$ \textit{ // Continue segmenting with SAM2.}
            \State $t \gets t+1$
        \EndIf
        \EndWhile    \\   
        \Return $\{\Bar{m}_1, ..., \Bar{m}_T\}$
    \end{algorithmic}
\end{algorithm}
\section{Experiments \& Results}
\label{sec:exp}

We start this section by describing the datasets used in our evaluations.
Subsequently, we describe the experimental setup used to train the models.
We then present frame-wise segmentation results before evaluating the temporal smoothness of video segmentation and eventually giving an overview of the large-scale annotations we derive from our method and make available to the general public.

\subsection{Datasets}

The \textbf{\emph{Cholec80}} dataset \cite{twinanda2016endonet} consists of 80 videos of laparoscopic cholecystectomy performed by 13 surgeons. The videos have an average length of $2,306.27$ seconds, are recorded at 25 FPS, and have a resolution of $854 \times 480$ or $1920 \times 1080$ pixels. They are annotated with one of seven surgical phases for each frame and multi-class multi-label annotations for seven surgical tools at 1 FPS.

Derived from \emph{Cholec80}, the \textbf{\emph{CholeSeg8k}} dataset \cite{hong2020cholecseg8k} contains 8080 frames of laparoscopic cholecystectomy, fully annotated with segmentation masks for 13 semantic labels, including black background, abdominal wall, liver, gastrointestinal tract, fat, grasper, connective tissue, blood, cystic duct, L-hook electrocautery, gallbladder, hepatic vein, and liver ligament.

The \textbf{\emph{CATARACTS}} challenge data \cite{al2019cataracts} was initially introduced as a challenge on surgical tool usage recognition and later on for surgical phase prediction. It consists of 50 video sequences of cataract surgery at 30 FPS, a $1920 \times 1080$ pixels resolution and an average length of $656.29$ seconds. Two experts annotated the tool usage of 21 surgical instruments. 

Introduced as a sub-challenge on semantic segmentation of cataract surgery images, the \textbf{\emph{CaDISv2}} dataset \cite{grammatikopoulou2021cadis} contains 4670 images of the 25 \emph{CATARACTS} training videos, which are fully annotated with segmentation masks. The total count of labels is 36, from which 28 are surgical instruments, four are anatomy classes, and three are miscellaneous objects appearing during the surgery. Our experiments focus on the pre-defined experiment setting II, which groups the instrument classes into ten classes, resulting in 17 semantic labels. 

Lastly, the \textbf{\emph{Cataract-1k}} dataset \cite{ghamsarian2023cataract} consists of over 1000 cataract surgery videos recorded at 60 FPS, from which different subsets are annotated for different tasks, including surgical phase prediction, semantic segmentation and irregularity detection. Here, we focus on the 30 videos from which 2256 frames are annotated with segmentation masks for the surgical instrument, pupil, iris and artificial lens. These frames have a resolution of $512 \times 384$ pixels. 

An analysis of the scarcity of annotations of the respective datasets can be found in Supplementary Section \ref{sec:app_scarce}.

\subsection{Experimental Setup}

We split the available videos in \emph{CholecSeg8k}, \emph{CaDISv2} and \emph{Cataracts1k} for training/validation/testing by 14/2/2, 19/3/3 and 24/3/3, respectively.
Our \emph{Overseer} models are trained for $1\text{e}5$ steps on the small-scale datasets with a batch size of 8. We are using the \emph{AdamW} optimiser \cite{loshchilov2017decoupled} with $(\beta_1=0.5,\beta_2=0.999)$, an initial learning rate of $1\text{e-}4$ and a weight decay of $0.05$. The learning rate is decayed every $2\text{e}4$ steps by a factor of $0.5$. To match the training configurations of the involved backbones, we rescale images to $(299 \times 299)$ pixels for \emph{Mask R-CNN} and \emph{Mask2Former} and $(200 \times 200)$ pixels for \emph{DETR}. The models have been trained on a single Nvidia RTX4090 using PyTorch 2.4.1 and Cuda 12.2.
Further details on the model and training configurations and the code to reproduce our results can be found at \href{https://github.com/MECLabTUDA/SASVi}{https://github.com/MECLabTUDA/SASVi} upon acceptance.

\subsection{Per-Frame Object Detection \& Segmentation Results}
\label{sec:overseer}

This section presents object detection and segmentation results on the small-scale annotated sub-datasets. For \emph{quantitative evaluation} of the bounding boxes, we deploy the IoU metric at a $50\%$ threshold. To evaluate the predicted classes of objects, we use the F1 score at a $50\%$ IoU threshold, and to quantify the per-object segmentation quality, we deploy the Dice metric at $50\%$ IoU. We additionally evaluate the final semantic segmentation quality using the macro-average Dice metric (\emph{Semantic Dice}).

\begin{figure}[htbp]
    \centering
    \if\usepng1
        \includegraphics[width=0.95\textwidth]{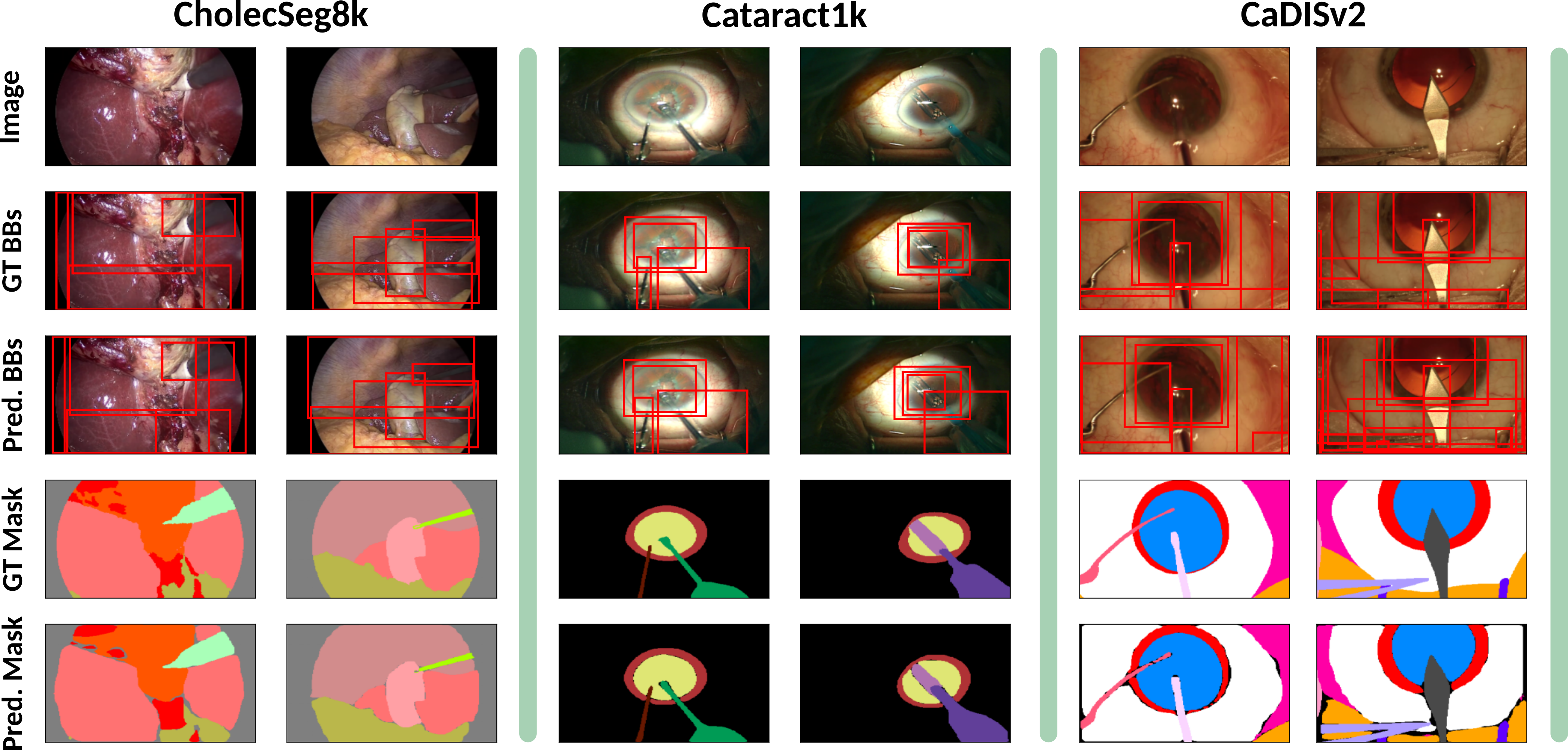}
    \else
        \includegraphics[width=0.95\textwidth]{figures/frame_wise_qual.eps}
    \fi
    \caption{\textbf{Qualitative Object Detection \& Segmentation Results.} Object detection methods such as \emph{Mask R-CNN} can serve as a powerful frame-wise \emph{Overseer} model, predicting classes, bounding boxes and segmentation masks of objects in surgical scenes.}
    \label{fig:frame_qual}
\end{figure}

The results of all metrics are displayed in Table \ref{tab:framewise}, and qualitative results for \emph{Mask R-CNN} are shown in Figure \ref{fig:frame_qual}. While \emph{Mask R-CNN} occasionally predicts multiple bounding boxes for the same object, resulting in lower per-object scores, it generally performs well across all datasets, especially regarding the final segmentation masks obtained.
However, the Transformer-based methods \emph{DETR} and \emph{Mask2Former} suffer less from this issue and generally show superior performance. We therefore opt to continue with \emph{Mask2Former} as our main \emph{Overseer} model for \emph{SAM2}

\begin{table}[htbp]
    \centering
    \resizebox{\textwidth}{!}{
        \begin{tabular}{c|c|cccc}
            \textbf{Dataset} & \textbf{Method} & \textbf{Class F1 $(\uparrow)$} & \textbf{BB IoU $(\uparrow)$}  & \textbf{Mask Dice $(\uparrow)$} & \textbf{Semantic Dice $(\uparrow)$} \\
            \hline
            & Mask R-CNN & 0.957 & 0.887 & 0.834 & 0.937 \\
            CholecSeg8k & DETR & 0.935 & \textbf{0.893} & 0.912 & 0.934 \\
            & Mask2Former & \textbf{0.958} & 0.884 & \textbf{0.913} & \textbf{0.940} \\
            \hline
            & Mask R-CNN & 0.585 & 0.636 & 0.626 & 0.786 \\
            CaDISv2 & DETR & 0.769 & 0.774 & 0.811 & \textbf{0.854} \\
            & Mask2Former & \textbf{0.823} & \textbf{0.824} & \textbf{0.828} & 0.838 \\
            \hline
            & Mask R-CNN & 0.745 & 0.731 & 0.664 & 0.881 \\
            Cataract1k Segm. & DETR & \textbf{0.835} & \textbf{0.777} & \textbf{0.777} & \textbf{0.897} \\
            & Mask2Former & 0.764 & 0.729 & 0.737 & 0.881 \\
        \end{tabular}
    }
    \caption{\textbf{Per-Frame Overseer Object Detection \& Segmentation Results.}}
    \label{tab:framewise}
\end{table}

\subsection{Temporally Consistent Video Segmentation}
\label{sec:vid_seg}

Applying frame-wise models of any kind onto sequential images often introduces artefacts of temporal inconsistencies due to ambiguities in predictions and a lack of temporal information \cite{rivoir2021long,frisch2023temporally}. Therefore, and due to the lack of large-scale ground truth annotations, we deploy the following metrics to quantify the quality and temporal consistency of video segmentations:
\begin{enumerate}
    \item Similarly to previous work on evaluating temporal consistency for image-to-image translation \cite{rivoir2021long,frisch2023temporally}, we deploy optical flow warping for evaluating the consistency of segmentations along the temporal axis. More specifically, given two subsequent image frames $v_t$ and $v_{t+1}$, we compute the optical flow $OF(v_t,v_{t+1})$ between them. We then use this optical flow in a warping operation $W$ to warp the previous segmentation mask as $m'_{t+1} := W(m_t,OF(v_t,v_{t+1}))$. We eventually compare the macro-average Dice and IoU scores of the warped segmentation $m'$ to the segmentation of the next frame $m_{t+1}$, denoted as $\text{Dice}_{OF}$ and $\text{IoU}_{OF}$ respectively.
    \item Analogously, we directly compute the macro-average Contour Distance and IoU scores of subsequent mask predictions $m_t$ and $m_{t+1}$, which we denote as $\text{CD}_{T}$ and $\text{IoU}_{T}$ respectively. Here, better scores indicate a better temporal consistency of the masks but disregard the actual image content.
\end{enumerate}
Appendix Section \ref{sec:app_temp} provides auxiliary visualisations for these metrics, and their results are presented in Table \ref{tab:videowise}. Qualitative results are presented in Figure \ref{fig:video_qual} with additional results in Section \ref{sec:app_qual} in the Appendix. For \emph{SAM2}, we prompt the model with the semantic mask predicted by \emph{Mask2Former} from the first frame (\emph{SAM2 ($t_1$)}). Further, we experiment with re-prompting the model with ground truth segmentation masks every time they are available, denoted as $\emph{SAM2 (GT)}$. We additionally compare the approaches to a frame-wise \emph{nnUNet} with the \emph{ResNetEncM} configuration \cite{isensee2021nnu}, trained on $(128 \times 128)$ sized images and an equal number of steps as the \emph{Overseer} models, and to Surgical De-SAM \cite{sheng2024surgical}, trained on $(1024 \times 1024)$ images until convergence.

\begin{table}[htbp]
    \centering
    \caption{\textbf{Quantitative Video Segmentation Results.}}
    \begin{tabular}{c|c|cccc}
         \textbf{Dataset} & \textbf{Method} & \textbf{$\text{Dice}_{OF} (\uparrow)$} & \textbf{$\text{IoU}_{OF} (\uparrow)$} & \textbf{$\text{CD}_{T} (\downarrow)$} & \textbf{$\text{IoU}_{T} (\uparrow)$} \\
         \hline
          & nnUNet & 0.562 & 0.476 & 6.811 & 0.573 \\
          & Mask R-CNN & 0.568 & 0.482 & 7.002 & 0.555 \\
          & Mask2Former & 0.625 & 0.542 & 4.654 & 0.624 \\
          & Surgical-DeSAM & 0.540 & 0.459 & 7.390 & 0.546 \\
         Cholec80 & SAM2 ($t_1$) & 0.451 & 0.398 & 163.98 & 0.475 \\
          & SAM2 (GT) & 0.730 & 0.636 & \textbf{2.879} & 0.769 \\
          & SASVi (Mask R-CNN) & 0.737 & 0.645 & 3.449 & 0.763 \\ 
          & SASVi (Mask2Former) & \textbf{0.754} & \textbf{0.662} & 3.291 & \textbf{0.780} \\
         \hline
          & nnUNet & 0.547 & 0.474 & 5.116 & 0.583 \\
          & Mask R-CNN & 0.375 & 0.308 & 6.134 & 0.501 \\
          & Mask2Former & 0.592 & 0.515 & 3.601 & 0.623 \\
          & Surgical-DeSAM & 0.518 & 0.437 & 4.621 & 0.560 \\
         CATARACTS & SAM2 ($t_1$) & 0.465 & 0.412 & 126.05 & 0.495 \\
          & SAM2 (GT) & 0.652 & 0.568 & \textbf{2.939} & 0.695 \\
          & SASVi (Mask R-CNN)& 0.658 & 0.570 & 3.466 & 0.694 \\ 
          & SASVi (Mask2Former) & \textbf{0.674} & \textbf{0.588} & 3.028 & \textbf{0.715} \\
         \hline
          & nnUNet & 0.662 & 0.570 & 1.951 & 0.690 \\
          & Mask R-CNN & 0.578 & 0.500 & 2.717 & 0.605 \\
          & Mask2Former & 0.665 & 0.575 & \textbf{1.911} & 0.681 \\
          & Surgical-DeSAM & 0.665 & 0.575 & 2.094 & 0.619 \\
         Cataract1k & SAM2 ($t_1$) & 0.329 & 0.292 & 241.53 & 0.339 \\
          & SAM2 (GT) & 0.726 & 0.630 & 1.980 & 0.744 \\
          & SASVi (Mask R-CNN) & \textbf{0.741} & \textbf{0.650} & 1.935 & \textbf{0.756} \\ 
          & SASVi (Mask2Former) & 0.730 & 0.634 & 1.986 & 0.751 \\
    \end{tabular}
    \label{tab:videowise}
\end{table}

\begin{figure}[htbp]
    \centering
    \if\usepng1
        \includegraphics[width=\textwidth]{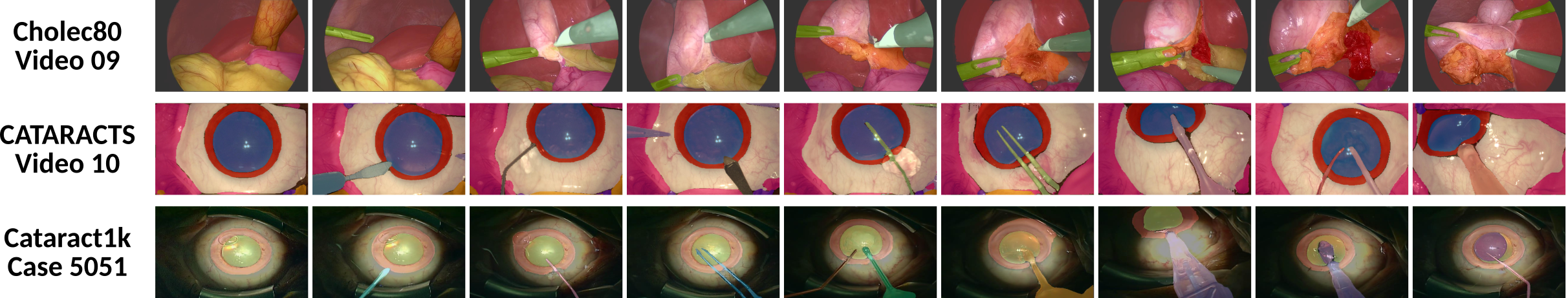}
    \else
        \includegraphics[width=\textwidth]{figures/video_qual.eps}
    \fi
    \caption{\textbf{Qualitative Video Segmentation Results.} \emph{SASVi (Mask R-CNN)} predicts smooth and complete annotations for surgical videos of arbitrary domains, here demonstrated for one video of \emph{Cholec80} (top), \emph{CATARACTS} (middle) and Cataract1k (bottom).}
    \label{fig:video_qual}
\end{figure}

Clearly, the re-prompting of \emph{SAM2}, be it from ground truth masks or our \emph{Overseer}, produces segmentations of significantly better temporal consistency. While \emph{SAM2 (GT)} predicts segmentations with lower \emph{Contour Distance} along the temporal axis, this can be explained by the metric's high sensitivity to outliers and not entirely optimal predictions from the \emph{Overseer}, as discussed in Section \ref{sec:overseer}. We are discussing this and other limitations and future improvements in Appendix Section \ref{sec:app_limit}. However, incorporating the actual image movement in the optical-flow-based metrics reveals better performance of \emph{SASVi} over all other considered methods. 

Our method allows us to leverage the \textbf{scarce annotations} available in \emph{CholecSeg8k}, \emph{CadISv2} and \emph{Cataract1k Segm.} and \textbf{produce full annotations of their large-scale video counterpart datasets} \emph{Cholec80}, \emph{CATARACTS} and \emph{Cataract1k}, respectively. Section \ref{sec:app_data} in the Appendix outlines the large-scale data statistics. We make those annotations available to the public, providing extensive annotation data for the future development of surgical analysis models.

\section{Conclusions}
\label{sec:concl}

We have presented \emph{SASVi}, a novel re-prompting mechanism for \emph{SAM2} based on a frame-wise object detection \emph{Overseer} model. Our novel contribution allows us to leverage the excellent temporal properties of \emph{SAM2} and smoothly and consistently segment arbitrary videos from various surgical domains with scarce annotation data. We have demonstrated the approach on three different surgical segmentation datasets covering cholecystectomy and cataract surgery. The obtained segmentation annotations for complete videos will be publicly available, enabling further development of surgical data science models and potentially mitigating class imbalance issues. We believe \emph{SASVi} can serve as a baseline for smooth and temporally consistent segmentation of surgical videos with scarcely available annotation data, taking surgical data science to the next level of automatisation.

\backmatter

\bmhead{Supplementary information}

The supplementary information comprises the Appendix of the main manuscript, including additional qualitative results in figure form and as video data. Additionally, we discuss limitations and future work and provide auxiliary visualisations for the temporal consistency metrics. Eventually, we also outline the data statistics for the large-scale annotations we generate by applying \emph{SASVi} to the full videos of the surgical datasets.






\section*{Declarations}

\textbf{Funding.} This work has been partially funded by the German Federal Ministry of Education and Research as part of the Software Campus programme (project 500 01 528).
\\
\textbf{Data Availability.} All experiments were conducted on publicly available datasets.
\\
\textbf{Code Availability.} Code will be published upon acceptance.
\\
Other declarations are not applicable.

\bibliography{sn-bibliography}

\newpage
\begin{appendices}

\section{Temporal Consistency Metrics}
\label{sec:app_temp}

\begin{figure}[htbp]
    \centering
    \if\usepng1
        \includegraphics[width=\textwidth]{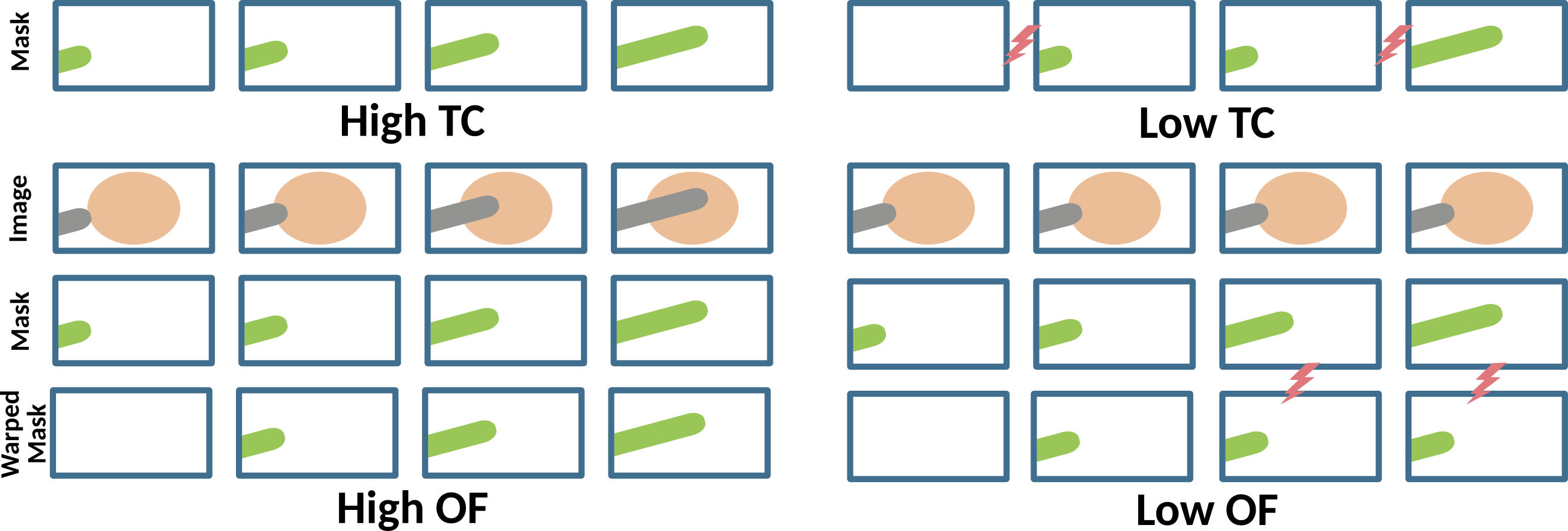}
    \else
        \includegraphics[width=\textwidth]{figures/tc_metrics.eps}
    \fi
    \caption{\textbf{Temporal Consistency Metrics.} The metrics $\text{CD}_{T}$ and $\text{IoU}_{T}$ consider the temporal consistency purely in mask space (top row). However, they fail to capture when images are stationary, but the masks transition smoothly.
    Therefore, $\text{Dice}_{OF}$ and $\text{IoU}_{OF}$ take the actual image movement into account, penalising such cases (bottom rows).}
    \label{fig:app_tc}
\end{figure}

This section aids in understanding the metrics introduced in Section \ref{sec:vid_seg} with simplified visualisations, displayed in Figure \ref{fig:app_tc}.

\section{Additional Qualitative Results}
\label{sec:app_qual}

\begin{figure}[htbp]
    \centering
    \includegraphics[width=\linewidth]{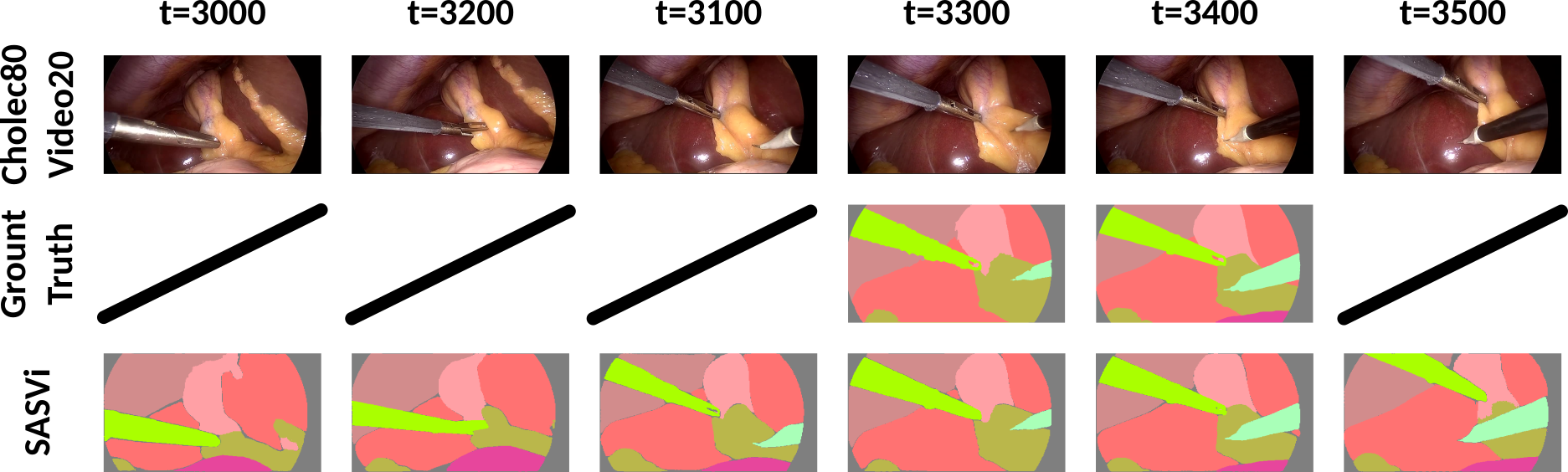}
    \caption{\textbf{Additional Qualitative Results.} \emph{SASVi} predicts complete segmentation masks for whole videos (bottom row) only relying on scarcely available annotation data (middle row), here demonstrated for \emph{Video20} of the \emph{Cholec80} dataset (top row).}
    \label{fig:app_qual}
\end{figure}

This section presents additional qualitative results in Figure \ref{fig:app_qual}. Fully segmented example videos of each of the three datasets can be found at \href{https://hessenbox.tu-darmstadt.de/getlink/fiW6NMDLQ1z8oGsj1PD8Kc81/}{https://hessenbox.tu-darmstadt.de/getlink/fiW6NMDLQ1z8oGsj1PD8Kc81/}. In the videos, we also visually compare \emph{SASVi} to \emph{nnUNet}, a popular meta-learning framework for frame-wise segmentation of medical images.

\section{Limitations \& Future Work}
\label{sec:app_limit}

The performance of \emph{SASVi} naturally depends on the performance of the \emph{Overseer} model, as analysed in Table \ref{tab:sensitivity}. Hence, we will explore other model choices in future work, focusing primarily on models that can be effectively trained on scarcely available ground truth data. Additional techniques for reducing error propagation, such as incorporating model uncertainty estimates, also yield a promising direction for future research. During the late stages of preparing the manuscript, the authors of \emph{SAM2} \cite{ravi2024sam} provided the means to fine-tune the model on custom data, which we will include in the future. Further, we will explore including existing ground truth data during \emph{SASVi} inference. Despite these limitations, our proposed approach can be a strong baseline for smooth and temporally consistent segmentation. The method lets us publicly provide large-scale annotations of complete videos from scarcely available data, as presented in the next section. 

\begin{table}[htbp]
    \centering
    \resizebox{\textwidth}{!}{\begin{tabular}{c|c|cccc}
        \textbf{Overseer (Annotations)} & \textbf{Semantic Dice $(\uparrow)$} & \textbf{$\text{Dice}_{OF} (\uparrow)$} & \textbf{$\text{IoU}_{OF} (\uparrow)$} & \textbf{$\text{CD}_{T} (\downarrow)$} & \textbf{$\text{IoU}_{T} (\uparrow)$} \\
        \hline
        Mask R-CNN (100\%) & 0.881 & 0.741 & 0.650 & 1.935 & 0.756 \\
        Mask R-CNN (50\%) & 0.879 & 0.567 & 0.489 & 2.088 & 0.719 \\
        Mask R-CNN (10\%) & 0.855 & 0.473 & 0.405 & 2.453 & 0.655 \\
        Mask R-CNN (1\%) & 0.756 & 0.352 & 0.299 & 93.153 & 0.447 \\
    \end{tabular}}
    \caption{\textbf{Impact of Overseer Performance on SASVi.} The \emph{Overseer} is trained with fewer training samples to assess \emph{SASVi} performance under data scarcity constraints.}
    \label{tab:sensitivity}
\end{table}

\section{Compute Analysis}
\label{sec:app_compute}

This section analyses the applicability of the methods for real-time segmentation of surgical videos using a single Nvidia RTX4090. We provide their parameter count and FPS for \emph{Cholec80} in Table \ref{tab:efficiency}. The results show that \emph{SASVi} does not introduce a significant computational overhead over \emph{SAM2}, which stems from our choice of lightweight object detection \emph{Overseer} models. These models can monitor surgical scenes more efficiently than traditional surgical segmentation pipelines, such as \emph{nnUNet} \cite{isensee2021nnu}.

\begin{table}[htbp]
    \centering
    \begin{tabular}{c|cc}
         \textbf{Method} & \textbf{Number of Parameters} & \textbf{FPS} \\
         \hline
         nnUNet & $269.4 \times 10^6$ & 4.633 \\
         Mask R-CNN & $45.8 \times 10^6$ & 49.456 \\
         DETR & $42.8 \times 10^6$ & 51.361 \\
         Mask2Former & $106.8 \times 10^6$ & 25.974 \\
         \hline
         SAM2 (t1) & $224.4 \times 10^6$ & 8.064 \\ 
         SASVi (Mask2Former) & $331.2 \times 10^6$ & 6.680
    \end{tabular}
    \caption{\textbf{Model Compute Evaluation for \emph{Cholec80}.}}
    \label{tab:efficiency}
\end{table}

\section{Dataset Annotation Sparsity}
\label{sec:app_scarce}

\begin{figure}[htbp]
    \centering
    \includegraphics[width=1\textwidth]{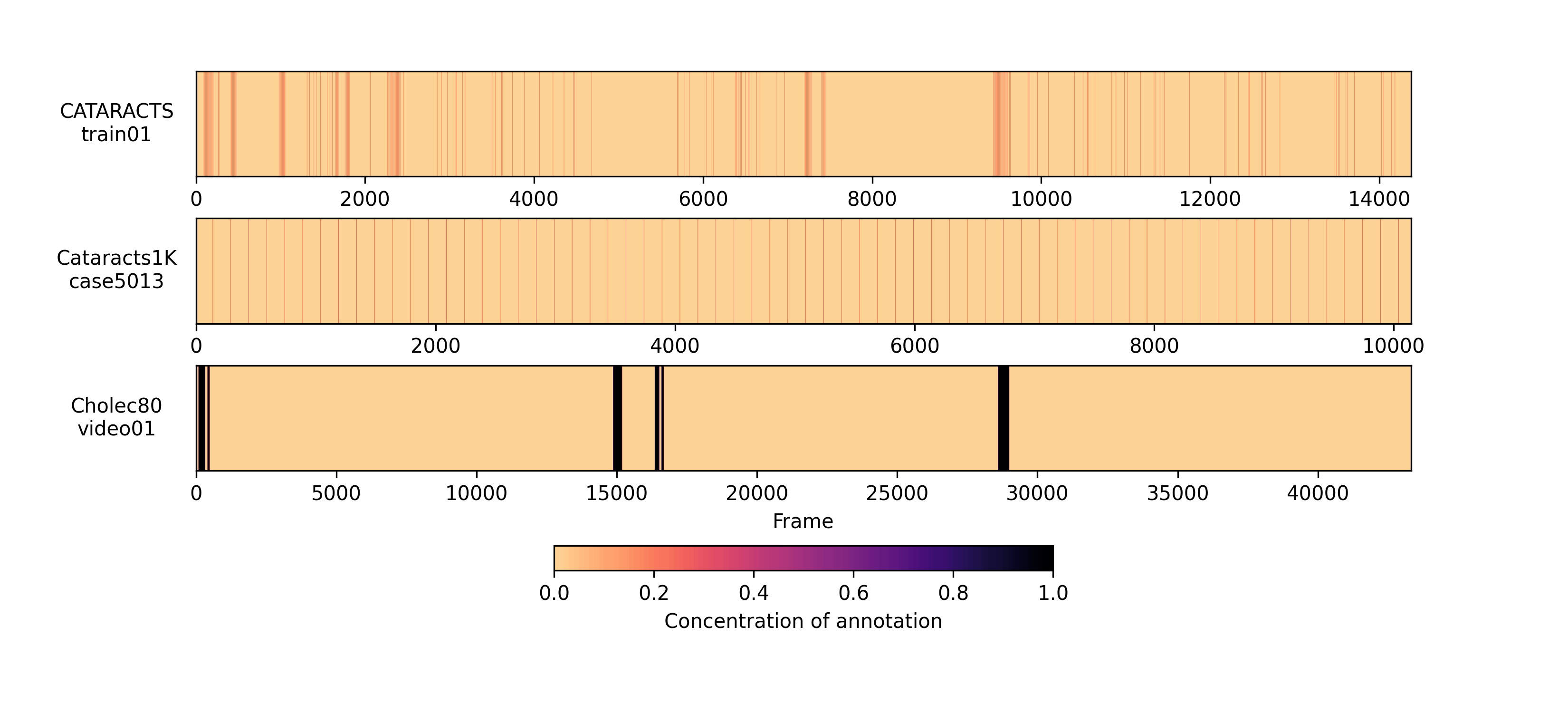}
    \caption{\textbf{Visualising Video Annotation Scarcity.} Each vertical bar represents one annotated frame. Multiple concentrated annotated frames blend into darker colours for visualisation.}
    \label{fig:temporal_plot}
\end{figure}

The three surgical datasets examined in this paper (\emph{CATARACTS} \cite{al2019cataracts}, \emph{Cataract1k} \cite{ghamsarian2023cataract} and \emph{Cholec80} \cite{twinanda2016endonet}) comprise full surgical videos each containing 50, 1000, and 80 videos respectively. We refer to these full videos as "large-scale datasets" or "counterparts". Each dataset only has a small subset of videos with only a few individual frames annotated with semantic segmentation masks: \emph{CaDISv2} \cite{grammatikopoulou2021cadis}, \emph{Cataract1k Segm.} \cite{ghamsarian2023cataract} and \emph{CholecSeg8k} \cite{hong2020cholecseg8k}, respectively. These annotations are scarce and vary significantly in length and distribution, as visualised in Figure \ref{fig:temporal_plot}. 

\begin{itemize}
\item \textbf{CATARACTS}: The videos were recorded at 30 FPS. Only 4670 out of 494,878 frames were annotated in the \emph{CaDISv2} subset \cite{grammatikopoulou2021cadis}, which constitutes just 0.95\% of the total frames. There are gaps as large as 5110 frames ($\approx$ 170 seconds) without annotations. 

\item \textbf{Cataract1k}: The videos were recorded at 60 FPS, with annotations provided at regular intervals of every 276th frame ($\approx$ 4.6 seconds) across 30 videos. This results in 2256 annotated frames, accounting for just 0.34\% of all available frames.

\item \textbf{Cholec80}: The videos were recorded at 25 FPS with an average length of 2306.27 seconds. While the \emph{CholeSeg8k} subset \cite{hong2020cholecseg8k} includes 8080 annotated frames, which is nearly twice as many as \emph{CaDISv2}, the annotations are only marginally denser, containing 1.08\% of annotated frames due to the videos being $\approx 3.5$ times longer on average. The annotations are also heavily concentrated at specific time frames, leaving extensive portions of the videos without any annotations.
\end{itemize}

The lack of datasets with continuous segmentation annotations in the surgical domain presents a significant challenge for training video segmentation models. Capturing temporal connections and modelling transitions across frames is difficult without such models. Hence, leveraging foundational models pre-trained on extensive and diverse datasets can help overcome this limitation by providing robust features for video segmentation in the surgical domain. 

\section{Large-Scale Annotation Data for Surgical Video Segmentation}
\label{sec:app_data}%

This section gives an overview of the large-scale annotations generated with \emph{SASVi} for the full video counterparts of the small-scale scarcely annotated data. Upon acceptance, we provide the obtained annotations for the public at \hyperlink{https://github.com/MECLabTUDA/SASVi}{https://github.com/MECLabTUDA/SASVi}, enabling future improvements of surgical data science models.

\begin{figure}[htbp]
    \centering
    \if\usepng1
        \includegraphics[width=\textwidth]{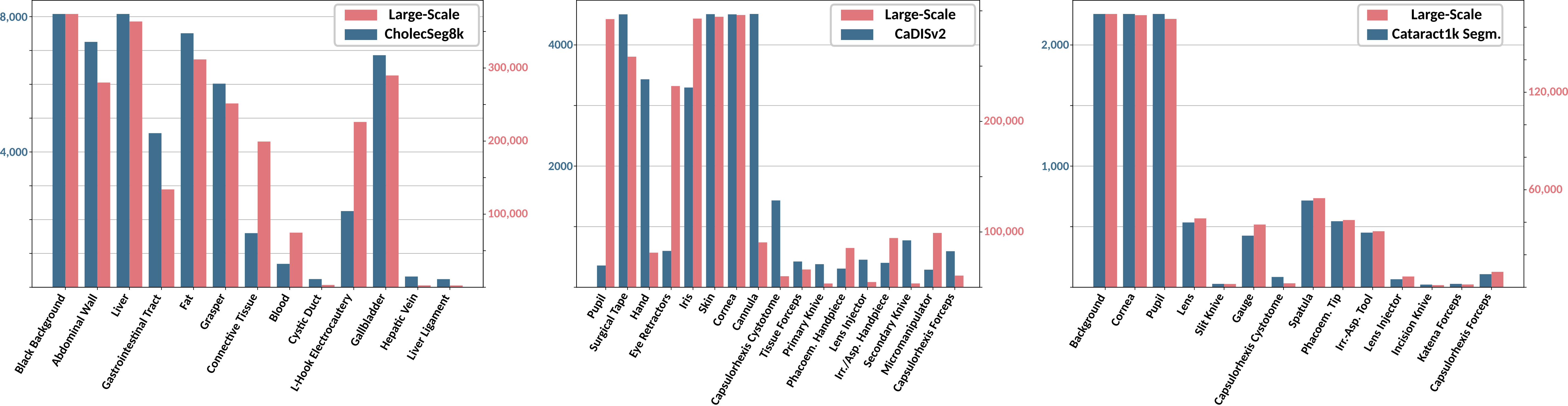}
    \else
        \includegraphics[width=\textwidth]{figures/data_stats.eps}
    \fi
    \caption{\textbf{Large-Scale Data Statistics.} Using \emph{SASVi}, we can greatly extend the available annotations for semantic segmentation of various surgical datasets, here demonstrated for \emph{Cholec80} (left), \emph{CATARACTS} (middle) and \emph{Cataract1k} (right). It is best viewed in the digital version.}
    \label{fig:app_data}
\end{figure}

We provide complete annotations for the 17 videos from \emph{Cholec80}, from which \emph{CholecSeg8k} was created. The left part of Figure \ref{fig:app_data} gives an overview of the available frames per label, comparing the previously available small-scale annotations and our large-scale extension. Analogously, we generate complete annotations for the 25 \emph{CATARACTS} videos from which the \emph{CaDIS} dataset was extracted. The middle part of Figure \ref{fig:app_data} displays the data statistics. Eventually, we also provide complete annotations for the 30 videos from which the \emph{Cataract1k} segmentation subset was extracted. The right part of figure \ref{fig:app_data} gives an overview of the statistics.

\end{appendices}

\end{document}